\documentstyle[erice98,12pt]{article}

\newcommand{\be}[1]{\begin{equation} \label{(#1)}}
\newcommand{\ee}{\end{equation}}
\newcommand{\ba}[1]{\begin{eqnarray} \label{(#1)}}
\newcommand{\ea}{\end{eqnarray}}

\newcommand{\G}{\gamma}
\newcommand{\GG}{{\gamma\gamma}}
\newcommand{\EPEM}{e^+e^-}

\newcommand{\A}{\alpha}

\newcommand{\BE}{\begin{equation}}
\newcommand{\EE}{\end{equation}}

\makeatletter
\def\lesssim{\mathrel{\mathpalette\vereq<}}
\def\vereq#1#2{\lower3pt\vbox{\baselineskip1.5pt \lineskip1.5pt
\ialign{$\m@th#1\hfill##\hfil$\crcr#2\crcr\sim\crcr}}}

\makeatother

\input{boxedeps.tex}
\SetRokickiEPSFSpecial
\HideDisplacementBoxes
\begin{document}
\title{Photon-Photon and Photon-Hadron Interactions at Relativistic
Heavy Ion Colliders}
\author{Gerhard Baur\footnote[1]{E-mail:G.Baur@fz-juelich.de}}
\address{address: Institut f\"ur Kernphysik,
Forschungszentrum J\"ulich, J\"ulich, Germany}
\author{Kai Hencken\footnote[2]{E-mail:hencken@quasar.physik.unibas.ch},
        Dirk Trautmann\footnote[3]{E-mail:trautmann@ubaclu.unibas.ch}}
\address{address: Institut f\"ur Physik,
Universit\"at Basel, Basel, Switzerland}

%
\maketitle

\abstracts{In central collisions at relativistic heavy ion colliders
like the Relativistic Heavy Ion Collider RHIC/Brookhaven and the Large
Hadron Collider LHC (in its heavy ion mode) at CERN/Geneva, one aims
at detecting a new form of hadronic matter --- the Quark Gluon Plasma.
We discuss here a complementary aspect of these collisions, the very
peripheral ones. Due to coherence, there are strong electromagnetic
fields of short duration in such collisions. They give rise to
photon-photon and photon-nucleus collisions with high flux up to an
invariant mass region hitherto unexplored experimentally. After a
general survey photon-photon luminosities in relativistic heavy ion
collisions are discussed. Then photon-photon physics at various
$\GG$-invariant mass scales is discussed.  The region of several GeV,
relevant for RHIC is dominated by QCD phenomena (meson and vector
meson pair production). Invariant masses of up to about 100 GeV can be
reached at LHC, and the potential for new physics is
discussed. Lepton-pair production, especially electron-positron pair
production is copious. Due to the strong fields there will be new
phenomena, especially multiple $\EPEM$ pair production.}

\section{Introduction}

Due to the coherent action of all the charges in the nucleus fast
nuclei are strong sources of equivalent (or quasireal) photons.  The
virtuality of the photon is related to the size $R$ of the nucleus by
\BE
Q^2 \lesssim 1/R^2, 
\EE
the condition for coherence. The maximum energy of the
quasireal photon is therefore given by
\BE
\omega_{max} \approx \frac{\G}{R},
\label{eq_wmax}
\EE
where $\gamma$ is the Lorentz factor.  We use natural units, setting
$\hbar=c=1$.

The collisions of $e^+$ and $e^-$ has been the traditional way to
study $\GG$-collisions. Similarly photon-photon collisions can also be
observed in hadron-hadron collisions. Since the photon number scales
with $Z^2$ ($Z$ being the charge number of the nucleus) such effects
can be particularly large. Of course, the strong interaction of the
two nuclei has to be taken into consideration.
%
%
\begin{figure}[tbhp]
\begin{center}
\ForceHeight{4cm}
\BoxedEPSF{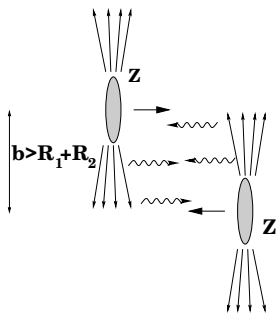}
\end{center}
\label{fig_collision}
\caption{\it
Two fast moving electrically charged objects are an abundant source of
(quasireal) photons. They can collide with each other and with the
other nucleus. For peripheral collisions with impact parameters $b>2R$,
this is useful for photon-photon as well as photon-nucleus collisions.}
\end{figure}

The equivalent photon flux present in medium and high energy nuclear
collisions is very high, and has found many useful applications in
nuclear physics \cite{BertulaniB88}, nuclear astrophysics
\cite{BaurR94,BaurR96}, particle physics \cite{Primakoff51} (sometimes
called the ``Primakoff effect''), as well as, atomic physics
\cite{Moshammer97}. With the construction of the ``Relativistic Heavy
Ion Collider'' (RHIC) and the ``Large Hadron Collider'' (LHC)
scheduled for 1999 and for 2004/2008, respectively, one will be able
to investigate such collisions experimentally. The main purpose of
these heavy ion colliders is the formation and detection of the
quark-gluon-plasma in central collisions. The present interest is in
the ``very peripheral (distant) collisions'', where the nuclei do not
interact strongly with each other. From this point of view, grazing
collisions and central collisions are considered as a background. It
is needless to say that this ``background'' can also be interesting
physics of its own.

The equivalent photon spectrum extends up to several GeV at RHIC
energies ($\G\approx 100$) and up to about 100 GeV at LHC energies
($\G\approx 3000$), see Eq.~(\ref{eq_wmax}).  Therefore the range of
invariant masses $M_{\GG}$ at RHIC will be up to about the mass of the
$\eta_c$, at LHC it will extend into an invariant mass range hitherto
unexplored. Up to now hadron-hadron collisions have not been used for
two-photon physics. An exception can be found in
\cite{Vannucci80}. There the production of $\mu^+\mu^-$ pairs at the
ISR was observed.  The special class of events was selected, where no
hadrons are seen associated with the muon pair in a large solid angle
vertex detector. In this way one makes sure that the hadrons do not
interact strongly with each other, i.e., one is dealing with
peripheral collisions (with impact parameters $b>2R$); the
photon-photon collisions manifest themselves as ``silent events''.
Dimuons with a very low sum of transverse momenta are also considered
as a luminosity monitor for the ATLAS detector at LHC
\cite{ShamovT98}.

Experiments are planned at RHIC
\cite{KleinS97a,KleinS97b,KleinS95a,KleinS95b,Nystrand98} and
discussed at LHC \cite{HenckenKKS96,Felix97,BaurHTS98}. We quote
J.D.Bjorken \cite{Bjorken99}: {\it It is an important portion (of the
FELIX program at LHC) to tag on Weizsaecker Williams photons (via the
nonobservation of completely undissociated forward ions) in ion-ion
running, creating a high luminosity $\gamma-\gamma$ collider.  }
Recent reviews are\cite{KraussGS97}, \cite{BaurHT98}, and
\cite{BaurHT98b}.

\section{From impact-parameter dependent equivalent photon spectra to
{$\GG$-luminosities}}
\label{sec_lum}

\begin{figure}[tbhp]
\begin{center}
\ForceHeight{4cm}
\BoxedEPSF{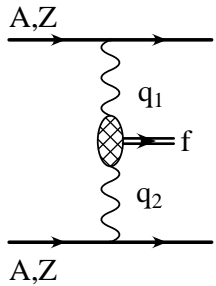}
\end{center}
\caption{\it
The general Feynman diagram of photon-photon processes in heavy ion
collisions: Two (virtual) photons fuse in a charged particle collision
into a final system $f$.
} 
\label{fig_ggcollision}
\end{figure}

Photon-photon collisions have been studied extensively at $\EPEM$
colliders. The theoretical framework is reviewed, e.g., in
\cite{BudnevGM75}.  The basic graph for the two-photon process in
ion-ion collisions is shown in Fig.~\ref{fig_ggcollision}. Two virtual
(space-like) photons collide to form a final state $f$. In the
equivalent photon approximation it is assumed that the square of the
4-momentum of the virtual photons is small, i.e., $q_1^2\approx
q_2^2\approx 0$ and the photons can be treated as quasireal. In this
case the $\GG$-production is factorized into an elementary cross
section for the process $\G+\G\rightarrow f$ (with real photons, i.e.,
$q^2=0$) and a $\GG$-luminosity function. In contrast to the pointlike
elementary electrons (positrons), nuclei are extended, strongly
interacting objects with internal structure. This gives rise to
modifications in the theoretical treatment of two photon processes.
The virtual photons in relativistic heavy ion collisions can be
treated as quasireal. This is a limitation as compared to $\EPEM$
collisions, where the two-photon processes can also be studied as a
function of the corresponding masses $q_1^2$ and $q_2^2$ of the
exchanged photon (``tagged mode'').
%
%

Relativistic heavy ions interact strongly when the impact parameter is
smaller than the sum of the radii of the two nuclei. In such cases
$\GG$-processes are still present and are a background that has to be
considered in central collisions \cite{Baur92,BaurB93b} . In order to
study ``clean'' photon-photon events however, they have to be
eliminated in the calculation of photon-photon luminosities as the
particle production due to the strong interaction dominates. In the
usual treatment of photon-photon processes in $\EPEM$ collisions plane
waves are used and there is no direct information on the impact
parameter. For heavy ion collisions on the other hand it is very
appropriate to introduce impact parameter dependent equivalent photon
numbers. They have been widely discussed in the literature (see, e.g.,
\cite{BertulaniB88,JacksonED,WintherA79}).  The cross section for a
certain electromagnetic process is then
\BE
\sigma = \int \frac{d\omega}{\omega} n(\omega) \sigma_{\G}(\omega).
\label{eq_sigmac}
\EE
A useful estimate is
\BE
n(\omega) \approx  \frac{2 Z^2 \A}{\pi} \ln
\frac{\G}{\omega R_{min}}.
\label{eq_nomegaapprox}
\EE

The photon-photon production cross-section is obtained in a similar
factorized form, by folding the corresponding equivalent photon
spectra of the two colliding heavy ions \cite{BaurF90,CahnJ90}
\BE
\sigma_c = \int \frac{d\omega_1}{\omega_1} \int
\frac{d\omega_2}{\omega_2}
F(\omega_1,\omega_2) \sigma_{\GG}(W_{\GG}=\sqrt{4 \omega_1\omega_2}) ,
\label{eq_sigmaAA}
\EE

\begin{figure}[htb]
\begin{center}
\ForceHeight{6cm}
\BoxedEPSF{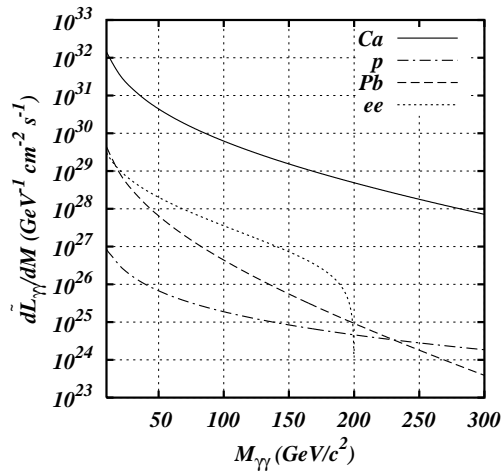}
\end{center}
\caption{\it
Comparison of the effective $\GG$-Luminosities ($L_{AA} \times
L_{\GG}$) for different ion species. For comparison the same quantity
is shown also for LEP200.
}
\label{fig_luminosity}
\end{figure}

In Fig.~\ref{fig_luminosity} we give a comparison of effective
$\gamma$$\gamma$ -luminosites (defined as collider luminosity times
$\GG$-luminosity) for various collider scenarios.
We use the following collider parameters: LEP200: $E_{el}=100$GeV,
$L=10^{32} cm^{-2} s^{-1}$, Pb-Pb heavy-ion mode at LHC: $\gamma$=2950,
$L=10^{26} cm^{-2} s^{-1}$, Ca-Ca: $\gamma$=3750,
$L=4 \times 10^{30} cm^{-2} s^{-1}$, p-p: $\gamma$=7450,
$L=10^{30} cm^{-2} s^{-1}$.
In the Ca-Ca heavy ion mode, higher effective luminosities can be
achieved as, e.g., in the Pb-Pb mode, since higher AA luminosities can
be reached there. For further details see \cite{HenckenTB95}.

\section{$\G$-A interactions}
\label{sec_ga}

There are many interesting phenomena ranging from the excitation of
discrete nuclear states, giant multipole resonances (especially the
giant dipole resonance), quasideuteron absorption, nucleon resonance
excitation to the nucleon continuum.  Photo-induced processes lead in
general to a change of the charge-to-mass ratio of the nuclei, and
with their large cross section they are therefore a serious source of
beam loss. Especially the cross section for the excitation of the
giant dipole resonance, a collective mode of the nucleus, is rather
large for the heavy systems (of the order of 100b).  The cross section
scales approximately with $Z^{10/3}$. Another serious source of beam
loss is the $\EPEM$ bound-free pair creation. The contribution of the
nucleon resonances (especially the $\Delta$ resonance) has also been
confirmed experimentally in fixed target experiments with 60 and~200
GeV/A (heavy ions at CERN, ``electromagnetic spallation'')
\cite{BrechtmannH88a,BrechtmannH88b,PriceGW88}. For details of these
aspects, we refer the reader to
\cite{KraussGS97,VidovicGS93,BaltzRW96,BaurB89}, where scaling laws,
as well as detailed calculations for individual cases are given.

The interaction of quasireal photons with protons has been studied
extensively at the electron-proton collider HERA (DESY, Hamburg), with
$\sqrt{s} = 300$~GeV ($E_e= 27.5$~GeV and $E_p=820$~GeV in the
laboratory system). This is made possible by the large flux of
quasi-real photons from the electron (positron) beam. The obtained $\G
p$ center-of-mass energies (up to $W_{\G p}\approx200$~GeV) are an
order of magnitude larger than those reached by fixed target
experiments. Similar and more detailed studies will be possible at the
relativistic heavy ion colliders RHIC and LHC, due to the larger flux
of quasireal photons from one of the colliding nuclei.  Estimates of
the order of magnitude of vector meson production in photon-nucleon
processes at RHIC and LHC are given in \cite{BaurHT98}.  At the LHC
one can extend these processes to much higher invariant masses $W$.
Whereas the $J/\Psi$ production at HERA was measured up to invariant
masses of $W\approx 160$~GeV, the energies at the LHC allow for
studies up to $\approx 1$~TeV.

At the LHC \cite{Felix97} hard diffractive vector meson
photoproduction can be investigated especially well in $AA$
collisions. In comparison to previous experiments, the very large
photon luminosity should allow observation of processes with quite
small $\G p$ cross sections, such as $\Upsilon$-production. For more
details see \cite{Felix97}.

\section{Photon-Photon Physics at various invariant mass scales}
\label{chap_proc}

Up to now photon-photon scattering has been mainly studied at $\EPEM$
colliders. Many reviews \cite{BudnevGM75,KolanoskiZ88,BergerW87} as
well as conference reports \cite{Sheffield95,Lund98} exist. The
traditional range of invariant masses has been the region of mesons,
ranging from $\pi^0$ ($m_{\pi^0}=135$~MeV) up to about $\eta_c$
($m_{\eta_c}=2980$~MeV).

The cross section for $\GG$-production in a heavy ion collision
factorizes into a $\GG$-luminosity function and a cross-section
$\sigma_{\GG}(W_{\GG})$ for the reaction of the (quasi)real photons
$\GG \rightarrow f$, where $f$ is any final state of interest.  When
the final state is a narrow resonance, the cross-section for its
production in two-photon collisions is given by
\BE
\sigma_{\GG\rightarrow R}(M^2) =
 8 \pi^2 (2 J_R+1) \Gamma_{\GG}(R) \delta(M^2-M_R^2)/M_R ,
\label{eq_nres}
\EE
where $J_R$, $M_R$ and $\Gamma_{\GG}(R)$ are the spin, mass and
two-photon width of the resonance $R$. This makes it easy to calculate
the production cross-section $\sigma_{AA\rightarrow AA+R}$ of a
particle in terms of its basic properties.  We will now give a general
discussion of possible photon-photon physics at relativistic heavy ion
colliders. Invariant masses up to several GeV can be reached at RHIC
and up to about 100 GeV at LHC.  An interesting topic in itself is the
$e^+$-$e^-$ pair production. The fields are strong enough to produce
multiple pairs in a single collisions. A discussion of this subjet
together with calculations within the semiclassical approximation can
be found in \cite{Baur90,HenckenTB95a,alscherHT97}

\subsection{Basic QCD phenomena in $\GG$-collisions}
\subsubsection{Light and heavy quark spectroscopy}

One may say that photon-photon collisions provide an independent view
of the meson and baryon spectroscopy. They provide powerful
information on both the flavor and spin/angular momentum internal
structure of the mesons. Much has already been done at $\EPEM$
colliders. Light quark spectroscopy is very well possible at RHIC,
benefiting from the high $\GG$-luminosities. Detailed feasibility
studies exist
\cite{KleinS97a,KleinS97b,KleinS95a,KleinS95b,Nystrand98}.  In this
study, $\GG$ signals and backgrounds from grazing nuclear and beam gas
collisions were simulated with both the FRITIOF and VENUS Monte Carlo
codes. The narrow $p_\perp$-spectra of the $\GG$-signals provide a
good discrimination against the background. The possibilities of the
LHC are given in the FELIX LoI \cite{Felix97}.

The absence of meson production via $\GG$-fusion is also of great
interest for glueball search. The two-photon width of a resonance is a
probe of the charge of its constituents, so the magnitude of the
two-photon coupling can serve to distinguish quark dominated
resonances from glue-dominated resonances (``glueballs'').  In
$\GG$-collisions, a glueball can only be produced via the annihilation
of a $q\bar q$ pair into a pair of gluons, whereas a normal $q\bar
q$-meson can be produced directly.  In a recent reference
\cite{Godang97} results of the search for $f_J (2220)$ production in
two-photon interactions were presented. There a very small upper limit
for the product of $\Gamma_{\GG} B_{K_sK_s}$ was given, where $B_{K_s
K_s}$ denotes the branching fraction of its decay into $K_s K_s$.
From this it was concluded that this is a strong evidence that the
$f_J(2220)$ is a glueball.

For charmonium production, the two-photon width $\Gamma_{\GG}$ of
$\eta_c$ (2960 MeV, $J^{PC} = 0^{-+}$) is known from experiment. But
the two-photon widths of $P$-wave charmonium states have been measured
with only modest accuracy.  For RHIC the study of $\eta_c$ is a real
challenge \cite{KleinS97b}; the luminosities are falling and the
branching ratios to experimental interest ing channels are small.

$C=-1$ vector mesons can be produced in principle by the fusion of
three (or, less important, five, seven, \dots) equivalent photons. The
cross section scales with $Z^6$. It is smaller than the contributions
discussed above, even for nuclei with large $Z$.

\subsubsection{Vector-meson pair production. Total hadronic
cross-section} 

There are various mechanisms to produce hadrons in photon-photon
collisions. Photons can interact as point particles which produce
quark-antiquark pairs (jets) , which subsequently hadronize. Often a
quantum fluctuation transforms the photon into a vector meson
($\rho$,$\omega$,$\phi$, \dots) (VMD component) opening up all the
possibilities of hadronic interactions .  In hard scattering, the
structure of the photon can be resolved into quarks and
gluons. Leaving a spectator jet, the quarks and gluon contained in the
photon will take part in the interaction.  It is of great interest to
study the relative amounts of these components and their properties.

The L3 collaboration recently made a measurement of the total hadron
cross-section for photon-photon collisions in the interval $5 GeV <
W_{\GG} < 75 GeV$ \cite{L3:97}. It was found that the $\GG
\rightarrow$hadrons cross-section is consistent with the universal
Regge behavior of total hadronic cross-sections.  The production of
vector meson pairs can well be studied at RHIC with high statistics in
the GeV region \cite{KleinS97a}.  For the possibilities at LHC, we
refer the reader to \cite{Felix97} and \cite{BaurHTS98}, where also
experimental details and simulations are described.

\subsection{$\GG$-collisions as a tool for new physics}

The high flux of photons at relativistic heavy ion colliders offers
possibilities for the search of new physics. This includes the
discovery of the Higgs-boson in the $\GG$-production channel or new
physics beyond the standard model, like supersymmetry or
compositeness.  While the $\GG$ invariant masses, which will be reached
at RHIC, will mainly be useful to explore QCD at lower energies, the
$\GG$ invariant mass range at LHC --- up to about 100 GeV --- will
open up new possibilities.

A number of calculations have been made for a medium heavy standard
model Higgs \cite{DreesEZ89,MuellerS90,Papageorgiu95,Norbury90}. For
masses $m_H < 2 m_{W^\pm}$ the Higgs bosons decays dominantly into
$b\bar b$.Chances of finding the standard model Higgs in this case are
marginal \cite{BaurHT98}.

An alternative scenario with a light Higgs boson was, e.g., given in
\cite{ChoudhuryK97} in the framework of the ``general two Higgs
doublet model''. Such a model allows for a very light particle in the
few GeV region. With a mass of 10~GeV, the $\GG$-width is about 0.1
keV. The authors of \cite{ChoudhuryK97} proposed to look for such a
light neutral Higgs boson at the proposed low energy
$\GG$-collider. We want to point out that the LHC Ca-Ca heavy ion mode
would also be very suitable for such a search.  In
Refs. \cite{DreesGN94,OhnemusWZ94} $\GG$-processes at $pp$ colliders
(LHC) are studied. It is observed there that non-strongly interacting
supersymmetric particles (sleptons, charginos, neutralinos, and
charged Higgs bosons) are difficult to detect in hadronic collisions
at the LHC. The Drell-Yan and gg-fusion mechanisms yield low
production rates for such particles. Therefore the possibility of
producing such particles in $\GG$ interactions at hadron colliders is
examined. Since photons can be emitted from protons which do not break
up in the radiation process, clean events can be generated which
should compensate for the small number.  In \cite{DreesGN94} it was
pointed out that at the high luminosity of
$L=10^{34}$cm${}^{-2}$s${}^{-1}$ at the LHC ($pp$), one expects about
16 minimum bias events per bunch crossing. Even the elastic $\GG$
events will therefore not be free of hadronic debris. Clean elastic
events will be detectable at luminosities below
$10^{33}$cm${}^{-2}$s${}^{-1}$. This danger of ``overlapping events''
has also to be checked for the heavy ion runs, but it will be much
reduced due to the lower luminosities.

\section{Conclusion}

Basic properties of electromagnetic processes in very peripheral
hadron-hadron collisions are described. The method of equivalent
photons is a well established tool to describe these kind of
reactions. Reliable results of quasireal photon fluxes and
$\GG$-luminosities are available. Unlike electrons and positrons heavy
ions and protons are particles with an internal structure. Effects
arising from this structure are well under control. A problem, which
is difficult to judge quantitatively at the moment, is the influence
of strong interactions in grazing collisions, i.e., effects arising
from the nuclear stratosphere and Pomeron interactions.

The high photon fluxes open up possibilities for photon-photon as well
as photon-nucleus interaction studies up to energies hitherto
unexplored at the forthcoming colliders RHIC and LHC.  Interesting
physics can be explored at the high invariant $\GG$-masses, where
detecting new particles could be within range. Also very interesting
studies within the standard model, i.e., mainly QCD studies will be
possible. This ranges from the study of the total $\GG$-cross section
into hadronic final states up to invariant masses of about 100~GeV to
the spectroscopy of light and heavy mesons.


\end{document}